\pdfoutput=1

\documentclass[twocolumn,numbers,sort&compress]{svjour3}

\smartqed

\usepackage{graphicx}
\usepackage{amssymb}
\usepackage[mathscr]{eucal}
\usepackage{amsmath,amssymb,amscd}
\usepackage{natbib}
\usepackage{hyperref}
\usepackage{color}

\hypersetup{pdftitle = The title of my PDF, pdfauthor = My name, pdfsubject= The subject, pdfkeywords = keyword1 keyword2 keyword3}
\hypersetup{colorlinks = true, linkcolor = blue, anchorcolor = red, citecolor = blue, filecolor = red, pagecolor = red, urlcolor = blue}

\journalname{}

\def \mb{\mathbb}

\def \bf{\mathbf}

\def \R{\mb R}                 

\def \a{\alpha}         

\newcommand {\sG} {\mathscr{G}}
\newcommand {\sS} {\mathscr{S}}

\newcommand {\cH} {\mathcal{H}}

\newcommand {\cW} {\mathcal{W}}

\newcommand{\EQ}[1]{\begin{equation}\begin{split} #1 \end{split}\end{equation}}

\def \and{\mbox{and}}

\newcommand{\orcid}[1]{\href{https://orcid.org/#1}{\includegraphics[width=8pt]{orcid-ID.png}}}

\begin{document}

\title{Numerical investigation on the Hill's type lunar problem with homogeneous potential}

\author{Yanxia Deng$^{1}$, Slim Ibrahim$^{2}$ and Euaggelos E. Zotos$^{3}$}

\institute{
  Y. Deng \at
    $^1$School of Mathematics (Zhuhai), \\
    Sun Yat-sen University, Zhuhai, \\
    Guangdong, China \\
    email: \url{dengyx53@mail.sysu.edu.cn}
  \and
  S. Ibrahim \at
    $^2$Department of Mathematics and Statistics, \\
    University of Victoria, \\
    Victoria, BC, Canada \\
    email: \url{ibrahims@uvic.ca}
  \and
  E.E. Zotos \at
    $^3$Department of Physics, School of Science, \\
    Aristotle University of Thessaloniki, \\
    GR-541 24, Thessaloniki, Greece \\
    $^{*}$Corresponding's author email: \url{evzotos@physics.auth.gr}
}

\date{Received: - / Accepted: - / Published online: -}

\titlerunning{Numerical investigation on the Hill's type lunar problem with homogeneous potential}

\authorrunning{Y. Deng et al.}

\maketitle

\begin{abstract}

We consider the planar Hill's lunar problem with a homogeneous gravitational potential. The investigation of the system is twofold. First, the starting conditions of the trajectories are classified into three classes, that is bounded, escaping, and collisional. Second, we study the no-return property of the Lagrange point $L_2$ and we observe that the escaping trajectories are scattered exponentially. Moreover, it is seen that in the supercritical case, with $\a \geq 2$, the basin boundaries are smooth. On the other hand, in the subcritical case, with $\a = 1$ the boundaries between the different types of basins exhibit fractal properties.


\end{abstract}

\section{Introduction}
\label{intro}

The N-body problem has a long-standing history and there are tremendous work on this problem. We refer the readers to these textbooks \cite{AM08,MO17,MS95}. It is well-known that for more than two bodies the Newtonian N-body problem is not integrable and is chaotic. The problem has also been extended to the homogeneous gravitational field, with potential of the order $r^{-\a}$, where $r$ is the mutual distance and $\a>0$. It seems that most of the work are on the \emph{weak} potential case where $\a<2$, and the Newtonian gravitation ($\a=1$) is the paradigmatic case among the weak potential case. For \emph{strong} force where $\a\geq 2$, \cite{G75,MG81} are among the earlier work. Using the action functional and the variational principle, Gordon \cite{G75} showed the existence of periodic orbits for strong force conservative system. McGehee \cite{MG81} gave a full description of the profile and speed of collisional orbits for the strong force Kepler problem. The strong force case exhibits significantly different behavior from the weak potential case, as can be readily seen in the Kepler problem. In a recent work \cite{DI20}, the authors classified the N-body problem in terms of the singularity versus global existence. In particular, the strong potential case has the feature that the critical point satisfies an energy minimizing property which enables us to classify the dynamics nicely. Contrary to the weak force case, this energy variational property fails thus no such classifications are available, and it is well-know that chaotic behavior appears. In a more recent work \cite{DI19}, the authors gave a nice characterization of the Hill's Lunar problem with homogeneous potential.  It turns out that there are simple smooth boundaries that distinguish colliding orbits from global ones for $\a\geq2$ under some energy threshold, while for $\a<2$ there are no simple boundaries and indeed they seem to be fractal as suggested by the numerical results in this paper. The analytical methods used the idea of ground states, which were extensively studied in the nonlinear dispersive equation community, for example \cite{NaSc12, AkIb18}.

This numeric project is motivated by \cite{Z17}, in which the author characterized initial conditions for colliding, bounded, and escaping orbits respectively for the restricted three-body problem. In that paper, one of the primaries generates a Newtonian gravitational potential $r^{-1}$, and the other primary generates a potential $r^{-\a}$ where $1\leq \a<2$. Numerical results in \cite{Z17} suggest that the boundaries distinguishing colliding orbits and global orbits become less ``fractal'' as $\a$ approaches $2$. Note that the author in \cite{Z17} called $1<\a<2$ the ``strong'' potential, but in our paper they are still ``weak'' potential.  Among the global orbits, it is important to understand which orbits remain bounded and which are escaping. We refer the readers to \cite{Z17,BTS96} and the references therein for more information on escapes of Hamiltonian systems and chaotic scattering. It is our goal to investigate the initial conditions of the Hill's Problem in terms of colliding, bounded, and escaping orbits for all $\alpha>0$.

For the supercritical Hill's problem $(\alpha>2)$, the authors in \cite{DIN19} give a preliminary characterization of the dynamics for the global solutions. In particular, exponentially scattering is generic among the global orbits, i.e. $\sG\setminus\sS^X$ has no interior, where $\sG$ denotes the set of initial conditions leading to global solutions and $\sS^X$ are those leading to exponential scattering. That is, any global solution of the supercritical Hill's lunar problem which does not scatter exponentially is unstable in the sense that small perturbation leads either to the collision or the exponential scattering. In this paper, we will simulate some solutions to support this result.

Lastly, in \cite{DI19} the authors proved that there are no heteroclinic orbits between the two Lagrange points for $\alpha\geq2$. We mention that this is different from the Newtonian case, where there are homoclinic tangle between the two Lagrange points \cite{WBW05}. and conjectured that there there are no homoclinic orbits either. See the One-pass theorem, also known as the no-return property in that paper. We will check this ``no-return'' property for $\alpha\geq2$ numerically.

The structure of the paper is as follows: In section 2 we present in detail the properties of the mathematical model of the Hill's lunar problem. In section 3 we describe the computational methods we used for obtaining the classification of the trajectories, and present the numerical results about the colliding, bounded, and escaping orbits. In section 4 we study the no-return property and present some simulations about the scattering solutions. Our paper ends with Section 5 where the discussions and the conclusions of our research are given.

\section{Presentation of the main problem}
\label{prob}

The three-body problem is a prototypical case in celestial mechanics. The system Sun-Earth-Moon can be considered as a typical example of the three-body problem. Using heuristic assumptions about the relative size of various physical constants, Hill was able to give the equations for the motion of the moon as an approximation from the general three-body problem.  The Hill's lunar problem can be derived from the general three-body problem using symplectic scaling method \cite{DI19,MS82}. Using a uniform rotating coordinate system, the Earth can be viewed as fixed at the origin, and the Sun with infinity mass at the positive $x$-axis infinitely far away. The ratio of the two infinite quantities is taken so that the gravitational attraction of the sun on the moon is finite. The motion of the moon in this rotating frame is then give by the following equation, where $(x,y)$ describes the position of the moon.

In particular, the planar Hill's equation with homogenous gravitational potential is given by

\begin{equation}\label{eq:hlp}\begin{cases}\ddot{x}-2\dot{y}&=-V_x\\\ddot{y}+2\dot{x}&=-V_y,\end{cases}\end{equation}
where \begin{equation}V(x,y)=-\frac{\a+2}{2}x^2-\frac{\a+2}{r^\alpha},\quad r=\sqrt{x^2+y^2}, \quad \alpha>0\end{equation} is known as the effective potential. When $\a=1$, (\ref{eq:hlp}) is the Newtonian Hill's Lunar Problem; when $\a\geq 2$, we shall call $V$ the \emph{strong potential}.

This vector field is well-defined everywhere except at the origin $(0,0)$. By the existence and uniqueness theorem of ODE, given $q(0)=(x(0),y(0))\neq (0,0)$ and $\dot{q}(0)\in \R^2$, there exists a unique solution $q(t)$ defined on the interval $[0, T_\mathrm{max})$, where $T_\mathrm{max}$ is maximal.

\begin{definition}
If $T_\mathrm{max}<\infty$, then the solution is said to experience a singularity at $T_\mathrm{max}$; otherwise, we say the solution exists globally.
\end{definition}

\begin{definition}
Given a global solution $q(t)$, if $|q(t)|\to\infty$ as $t\to\infty$, we say the solution is scattering, moreover, if $|q(t)|\sim e^{kt}$ as $t\to\infty$ for some constant $k>0$, we say $q(t)$ is exponentially scattering.
\end{definition}

Since the ODE is locally Lipschitz in $r\neq0$, blow-up is possible only by approaching the unique singularity, namely the collision. In particular, for the Hill's equation, if $T_\mathrm{max}<\infty$, then \EQ{\lim\limits_{t\to T_\mathrm{max}} (x(t), y(t))= (0,0),} that is, the singularity of the Hill's equation is due to finite time collision at the origin.

The Hill's equation admits the famous Jacobi integral which we shall refer to as the energy, \begin{equation}E(x,y,\dot{x},\dot{y}):=\frac{1}{2}(\dot{x}^2+\dot{y}^2)+V(x,y).\end{equation}
The effective potential $V(x,y)$ has exactly two critical points $L_1:=(-\alpha^{\frac{1}{\alpha+2}},0)$ and $L_2:=(\alpha^{\frac{1}{\alpha+2}},0)$, which are known as the Lagrange points. We have that $\pm Q:=(\pm\alpha^{\frac{1}{\alpha+2}},0,0,0)$ are the only equilibria of (\ref{eq:hlp}). In \cite{DI19}, we define $\pm Q$ to be the \emph{ground states} as their energies are the minimum under some constraint . In particular, the \emph{ground state energy} $E^*$ is defined as \EQ{E^*:=\inf\{E| W=0\}}where \EQ{W:=-xV_x-yV_y=0.} It is shown that $E^*$ is exactly achieved by $\pm Q$. In particular, \EQ{E^*=E(\pm Q)=-\frac{1}{2}(\a+2)^2\a^{-\frac{\a}{\a+2}}.}

\begin{figure*}[!t]
\centering
\resizebox{\hsize}{!}{\includegraphics{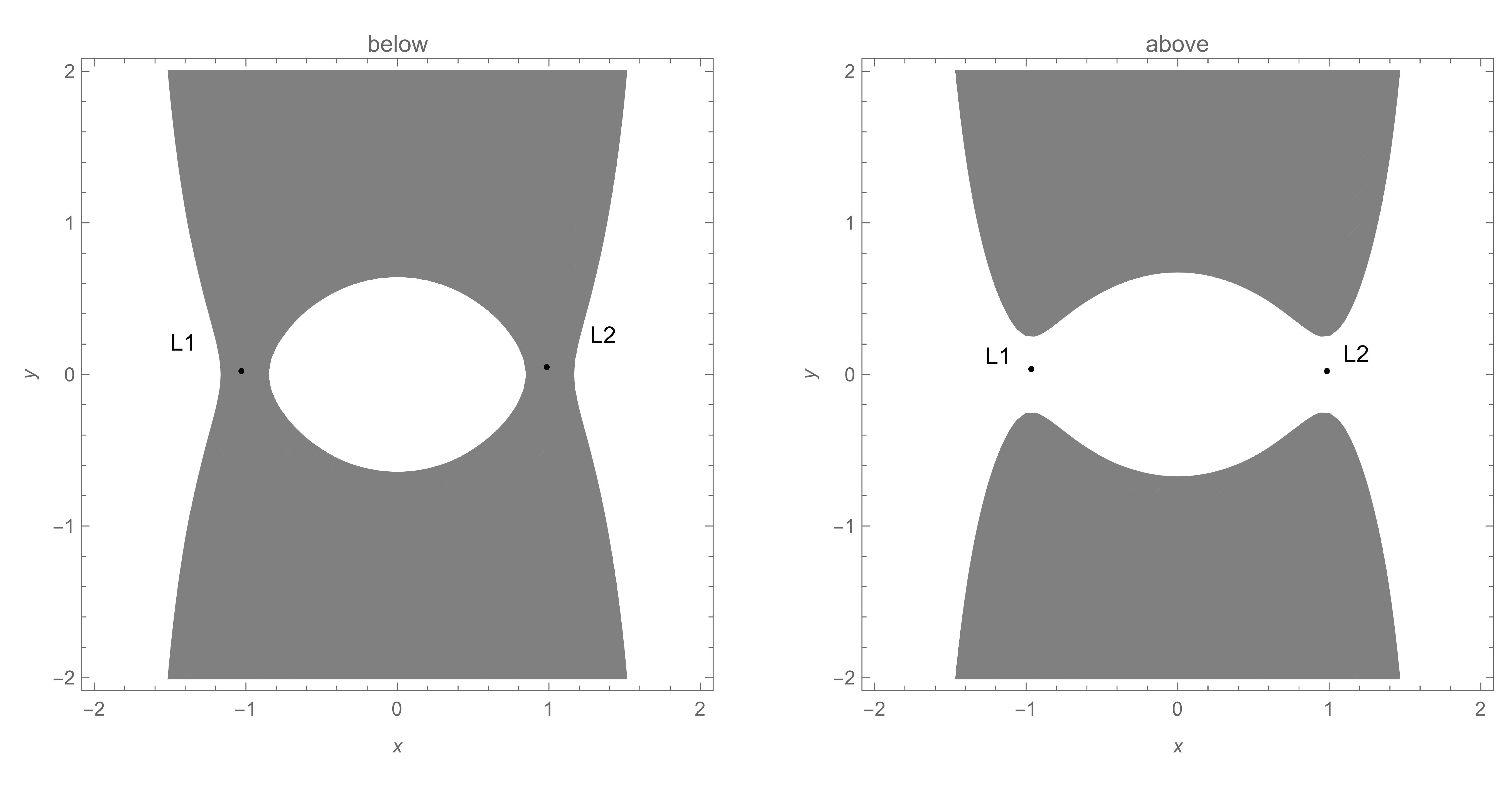}}
\caption{Hill's regions $\cH_c$ when $\a=1$, $E^*=-4.5$. The white regions correspond to the Hill's regions, and the gray shaded regions represent the energy forbidden regions. Left is below the ground state energy with $c=-4.6$; right is above the ground state energy with $c=-4.4$.}
\label{fig:Hillrg}
\end{figure*}

When one projects the four-dimensional phase space onto the configuration $(x,y)$ space, the resulting region is called the Hill's region.
\EQ{\cH_c:=\{(x,y) | E(x,y,\dot{x},\dot{y})=c\}=\{(x,y) | V(x,y)\leq c\}.}
The boundaries of the Hill's regions are called the \emph{zero velocity curves} because they are the position in the $(x,y)$-plane where the kinetic energy vanishes. The shapes of the Hill's regions rely on the values of the energy. There are four different cases in terms of the shape of the Hill's regions:
\begin{itemize}
\item $c<E^*$: both necks are closed, so orbits inside will remain bounded in the configuration space or collide with the origin.
\item $c=E^*$: the threshold case.
\item $E^*<c<0$: both necks are open, thus allowing orbits to enter the exterior region and escape from the system.
\item $c\geq 0$: motions over the entire configuration $(x,y)$ space is possible.
\end{itemize}

In Figure \ref{fig:Hillrg} we present the structure of the first and third possible Hill's region for $\a=1$; all the other $\a>0$ have the same structure with varied values of $L_1, L_2$.

A theorem that describes the fates of the solutions below the ground state energy $E^*$ is

\begin{theorem}[Dichotomy below the ground state \cite{DI19}]
\label{thm:below}
For the Hill's lunar problem, consider the sets: \begin{equation}
\begin{split}
\mathcal{W}_+&=\{\Gamma=(x, y, \dot{x}, \dot{y}) | E(\Gamma)<E^*, W(\Gamma) > 0\}\\
\mathcal{W}_-&=\{\Gamma=(x, y, \dot{x}, \dot{y}) | E(\Gamma)<E^*, W(\Gamma) \leq 0\}
\end{split}
\end{equation}
then $\mathcal{W}_+$ and $\mathcal{W}_-$ are invariant. Solutions in $\mathcal{W}_+$ exist globally and solutions in $\mathcal{W}_-$ are bounded or collide with the origin in finite time. Moreover, for $\a\geq2$, solutions in $\mathcal{W}_-$ all collide with the origin in finite time.
\end{theorem}

\begin{figure}[!t]
\centering
\resizebox{\hsize}{!}{\includegraphics{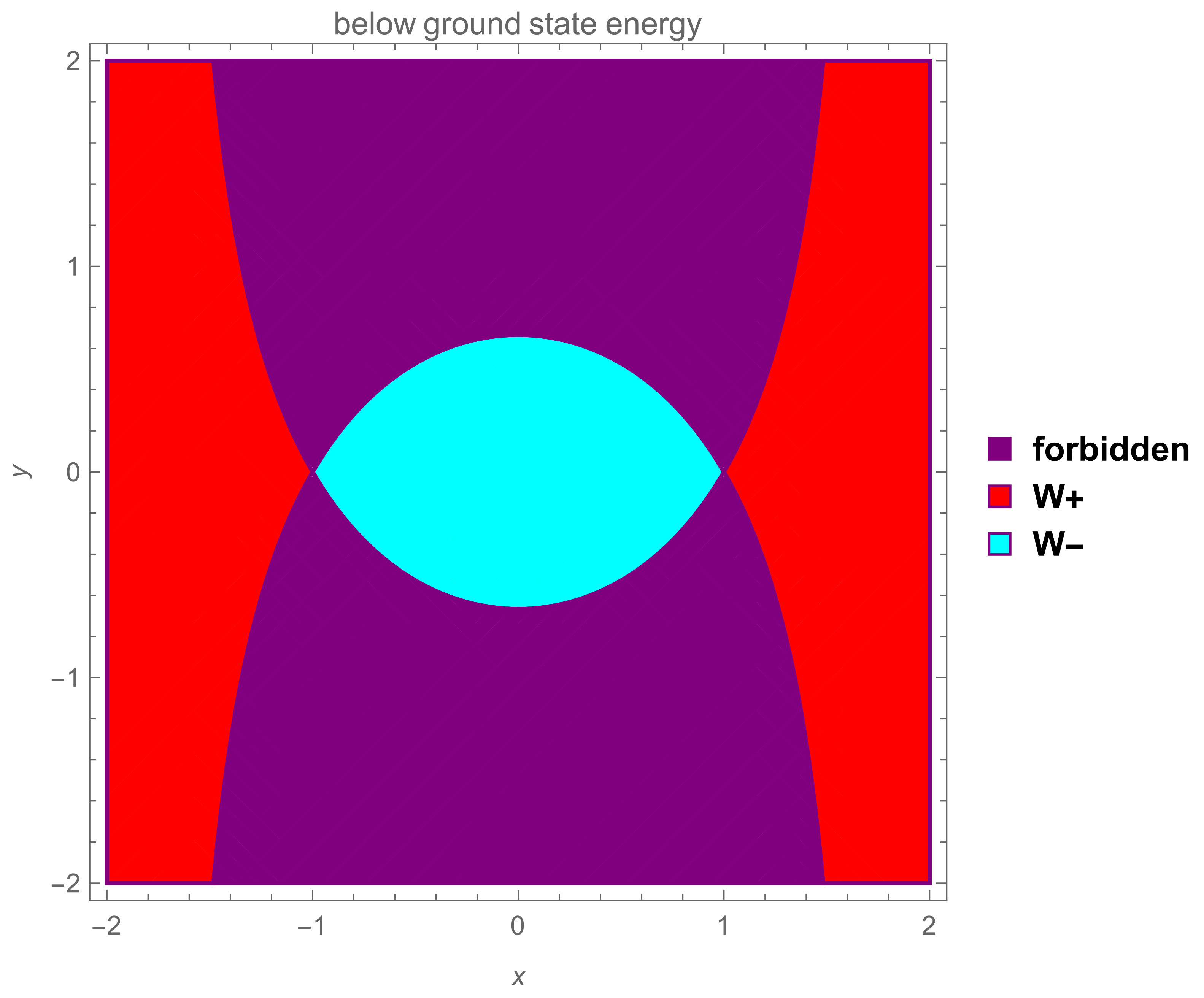}}
\caption{Projection of $\cW_\pm$ onto the configuration $(x,y)$ space. Plotted for $\a=1$, all other $\a>0$ have the similar structure.}
\label{fig:Kpm}
\end{figure}

Figure \ref{fig:Kpm} illustrates the region $\cW_\pm$ projected onto the configuration $(x,y)$ space, and all $\a>0$ has this structure. For $\a<2$, we don't necessarily have the finite time collision property in $\cW_-$, i.e. solutions in $\cW_-$ for $\a<2$ is either bounded globally or collides with the origin in finite time. Moreover, it is difficult to distinguish the bounded and the colliding orbits in $\cW_-$ for $\a<2$, as the boundaries of the region are fractal as we shall see in the following Figure \ref{fig:Orbelow}. The fractal geometry of the basin boundaries has also been confirmed by computing both the uncertainty dimension \cite{AVS01,AVS09} and the basin entropy \cite{DWGGS16}.

For energies equal to or above the ground state energy, there seem to be no simple sets to distinguish the fates based on the initial conditions. But still, we are able to describe the global dynamics for $\a\geq 2$, see \cite{DI19} and \cite{DIN19}. In this paper, we will simulate the no-return property for the supercritical case and show that the escaping orbits scatter exponentially as proved in \cite{DI19} and \cite{DIN19}.

\section{Orbit Classification}
\label{clas}

\begin{figure*}[!t]
\centering
\resizebox{0.7\hsize}{!}{\includegraphics{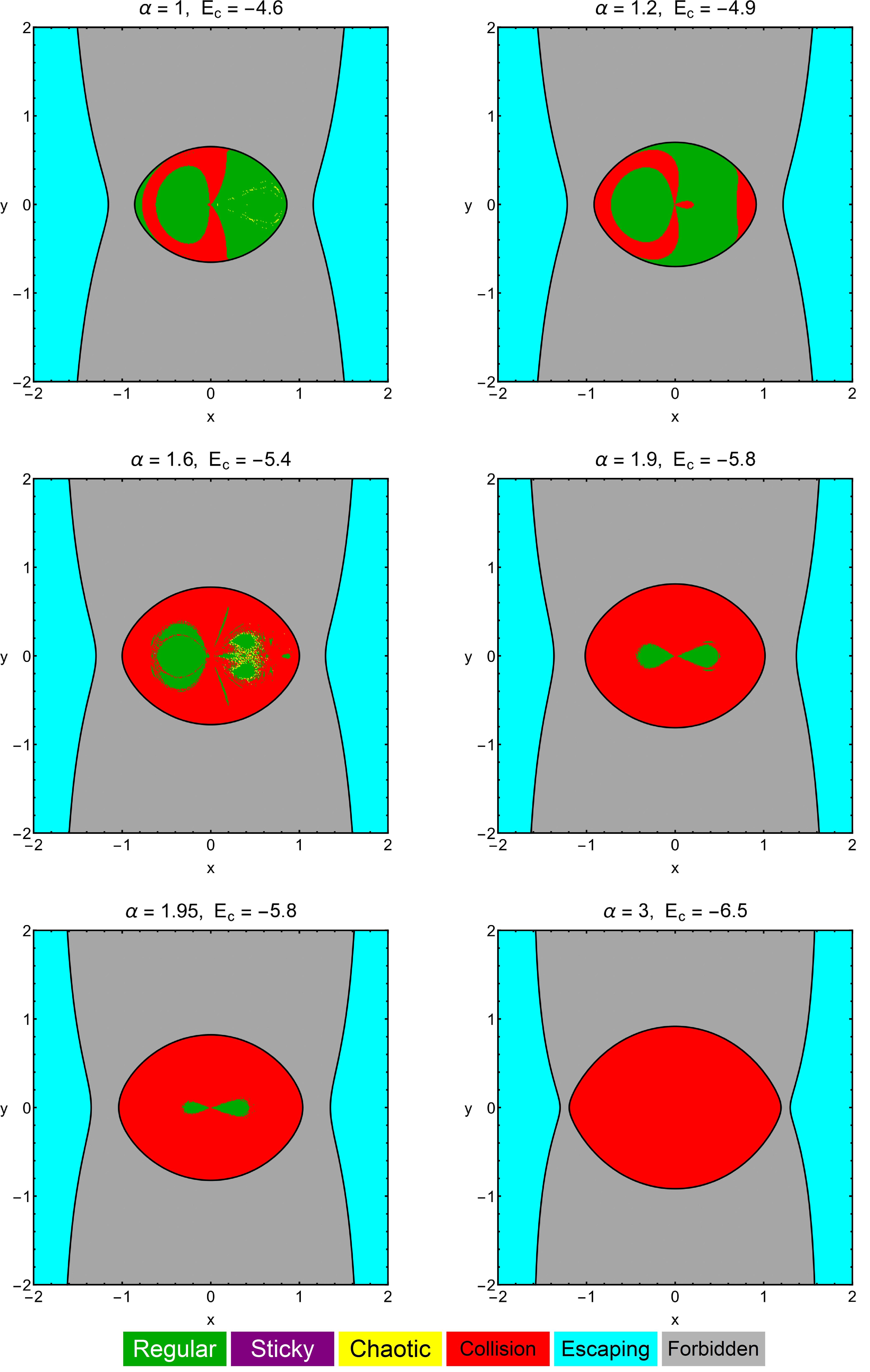}}
\caption{The orbital structure of the $\dot{y}>0$ part of the surface of section $\dot{x}=0$ \textbf{below} $E^*$. The values of $\alpha$ and $E_c$ are on top of each diagram.}
\label{fig:Orbelow}
\end{figure*}

\begin{figure*}[!t]
\centering
\resizebox{0.7\hsize}{!}{\includegraphics{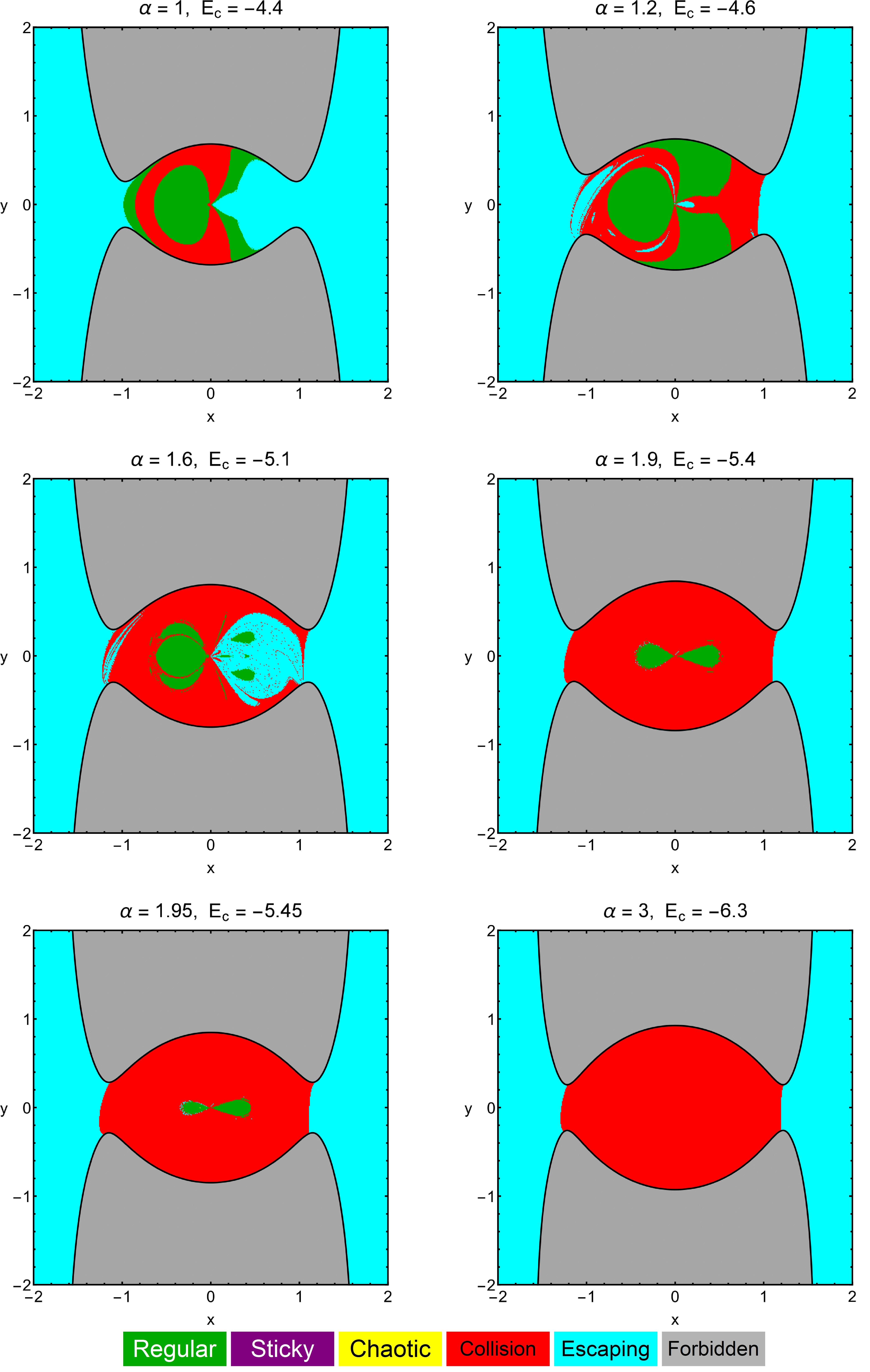}}
\caption{The orbital structure of the $\dot{y}>0$ part of the surface of section $\dot{x}=0$ \textbf{below} $E^*$. The values of $\alpha$ and $E_c$ are on top of each diagram.}
\label{fig:Orabove}
\end{figure*}

\begin{figure*}[!t]
\centering
\resizebox{0.7\hsize}{!}{\includegraphics{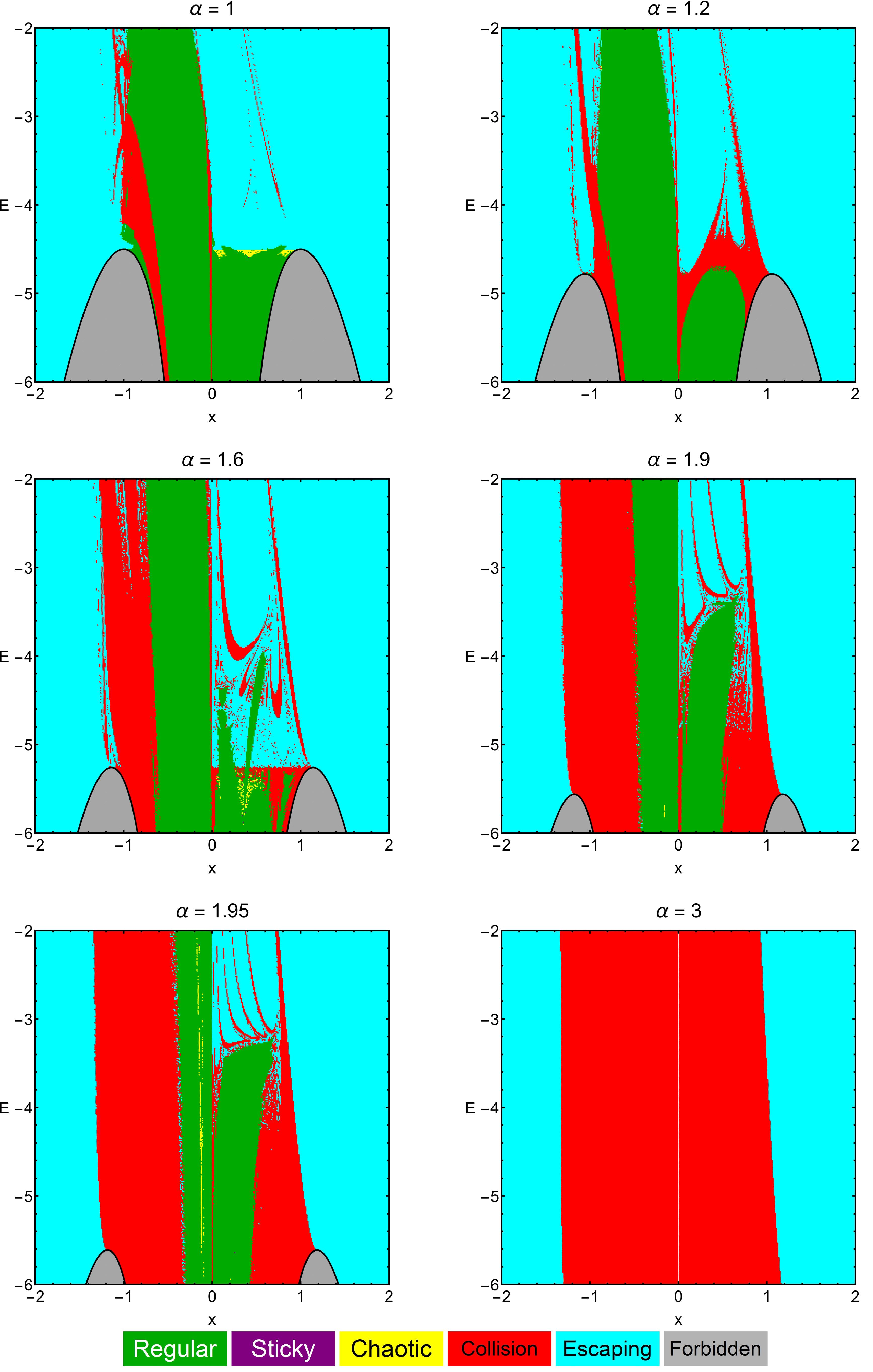}}
\caption{The orbital structure of the $(x,E)$ plane. The values of $\alpha$ are on top of each diagram.}
\label{fig:xE}
\end{figure*}

\subsection{Computational methods for Orbit Classification}

The motion of the infinitesimal test particle is restricted to a three-dimensional surface $E=E_c=const$. The condition $\dot{x}=0$ defines a two-dimensional surface of section, with two disjoint parts $\dot{y}<0$ and $\dot{y}>0$. Each of these two parts has a unique projection onto the $(x,y)$ configuration space, and we will take the part with $\dot{y}>0$. For each gravitational power $\alpha$ we take two values of the energy constant $E_c$, one below $E^*$ and one above $E^*$. For each fixed $\alpha$ and $E_c$, we define dense uniform grids of $1024\times 1024$ initial conditions regularly distributed on the configuration $(x,y)$ space inside the region allowed by the energy constant. The orbits are integrated with initial conditions inside a certain region, which in our case is a square grid with $-2\leq x, y\leq 2$.

In the Hill's problem, there are three possible types of motion for the test particle: (a) finite time collision into the origin; (b) bounded motion around the origin; (c) escape to infinity. We need to define appropriate numerical criteria to distinguish these three types of motion. The motion is considered bounded if the test particle stays inside the disk of radius $R_{\mathrm{esc}}$ centered at the origin for maximal integration time $t^*$. In our program, we take $t^*=5000$ time units and $R_{\mathrm{esc}}=10$. A trajectory is identified as escaping and the numerical integration stops if the test particle exits the disk of radius $R_{\mathrm{esc}}$ centered at the origin at a time $t\leq t^*$. Finally, the motion is considered as collision if the test particle crosses the disk with radius $R_{\mathrm{col}}$ around the origin. We choose $R_{\mathrm{col}}=10^{-3}$.

Note that we are aware of the fact that the higher the values of $R_{\mathrm{esc}}$ and $t^*$ the more plausible becomes the definition of bounded and escaping orbits. As a result, the higher these two values, the longer the numerical integration of initial conditions of the orbits lasts. However, the maximal numerical integration time $t^*=4$ is effective based on our theoretical results in \cite{DI19,DIN19}, and the vast majority of escaping orbits need considerable less time than $t^*$ to escape from the disk of radius $R_{\mathrm{esc}}$. We choose $R_{\mathrm{esc}}=10$ as suggested from \cite{Z17}. In \cite{Z17} and \cite{BTS96} the authors rely on the positivity of the total orbital energy measured by an observer in the inertial reference frame to validate that the escaping radius $R_{\mathrm{esc}}=10$ is safe. Though we didn't find the theoretical proof of the claim that positive orbital energy implies escaping, we point that our theoretical results in \cite{DI19}\cite{DIN19} show that solutions in $\cW_+$ escape to infinity, and our numerical results in Figure \ref{fig:Orbelow} match the theoretical prediction, hence imply that our choice of $R_{\mathrm{esc}}=10$ is valid. Nonetheless, we will check the positivity of the total orbital energy
\EQ{E_{\mathrm{orb}}:=E_{\mathrm{ki}}+E_{\mathrm{po}},}where \EQ{E_{\mathrm{ki}}:=\frac{1}{2}[(\dot{x}-y)^2+(\dot{y}+x)^2],\quad E_{\mathrm{po}}:=-\frac{1}{r^\a}} are the inertial kinetic energy and potential energy.

In the case of bounded motion, we apply the SALI chaos indicator \cite{S01} for further classifying between non-escaping regular orbits and trapped sticky and chaotic trajectories of the test particle. 

The equations of motion (\ref{eq:hlp}) for the initial conditions of all orbits are forwarded integrated using a Bulirsch-Stoer algorithm, with double-precision, developed in \verb!FORTRAN 77! \cite{PTVF92}. The energy integral of motion was conserved in $10^{-12}$. All graphical illustrations presented in this work have been created using Python version 3.7.3 and version 12.0 of the software Mathematica \cite{W03}.

The main numerical task is to classify initial conditions of orbits in the $\dot{y}>0$ part of the surface of section $\dot{x}=0$ into three categories. The initial conditions of orbits on the $(x,y)$ plane are classified into collision orbit, bounded orbit and escaping orbit. For each point in the $1024\times 1024$ grid with $-2\leq x, y\leq 2$, it is assigned a specific color, according to its final state.

In the following we will explore the orbital content of the configuration $(x,y)$ space in two different energy cases: one is below the ground state energy $E^*$, and the other is above $E^*$. In each case we choose six values of the power $\alpha$, namely $\alpha = 1.0, 1.2, 1.6, 1.9, 1.95, 3.0$. We choose these values to see the bifurcation from ``fractal'' to smooth boundaries between the colliding and global orbits when $\alpha\to 2^-$ as mentioned in the introduction. For every value of $\alpha$ the two energy levels are different, and they are taken roughly $E^*\pm0.15$. We call each of the color-coded grids an orbit type diagram - OTD as in \cite{N04,N05,Z17}.

\subsection{Below the ground state energy}

In this energy region both bottlenecks are closed (cf. Figure \ref{fig:Hillrg}), thus inside the interior region there are only collision and bounded orbits. In Figure \ref{fig:Orbelow} the OTD decomposition shows the orbital structure of the $(x,y)$-plane for six values of the gravitational power $\alpha$ and $E_c$ below the ground state energy $E^*$ for each $\alpha$. For each orbit that has been computed with initial position at the point $(x, y)$, we color that point with its orbital type. In Figure \ref{fig:Orbelow}, we see all points in the exterior region exhibit escaping motions. We are more interested in the interior region. In Figure \ref{fig:Orbelow} top left, where $\alpha=1$ (classical Newtonian gravity), we see that inside the interior region, both collision motion and bound motion are presented. In the other diagrams of Figure \ref{fig:Orbelow}, as $\alpha$ increases, the collision basins increase and fill the entire interior region when $\alpha=2$ and beyond. This serves as a numerical illustration for Theorem \ref{thm:below}. Moreover, the boundaries for collision and bounded orbits become smoother with respect to that observed in diagram $\alpha=1$.

\begin{figure*}[!t]
\centering
\resizebox{\hsize}{!}{\includegraphics{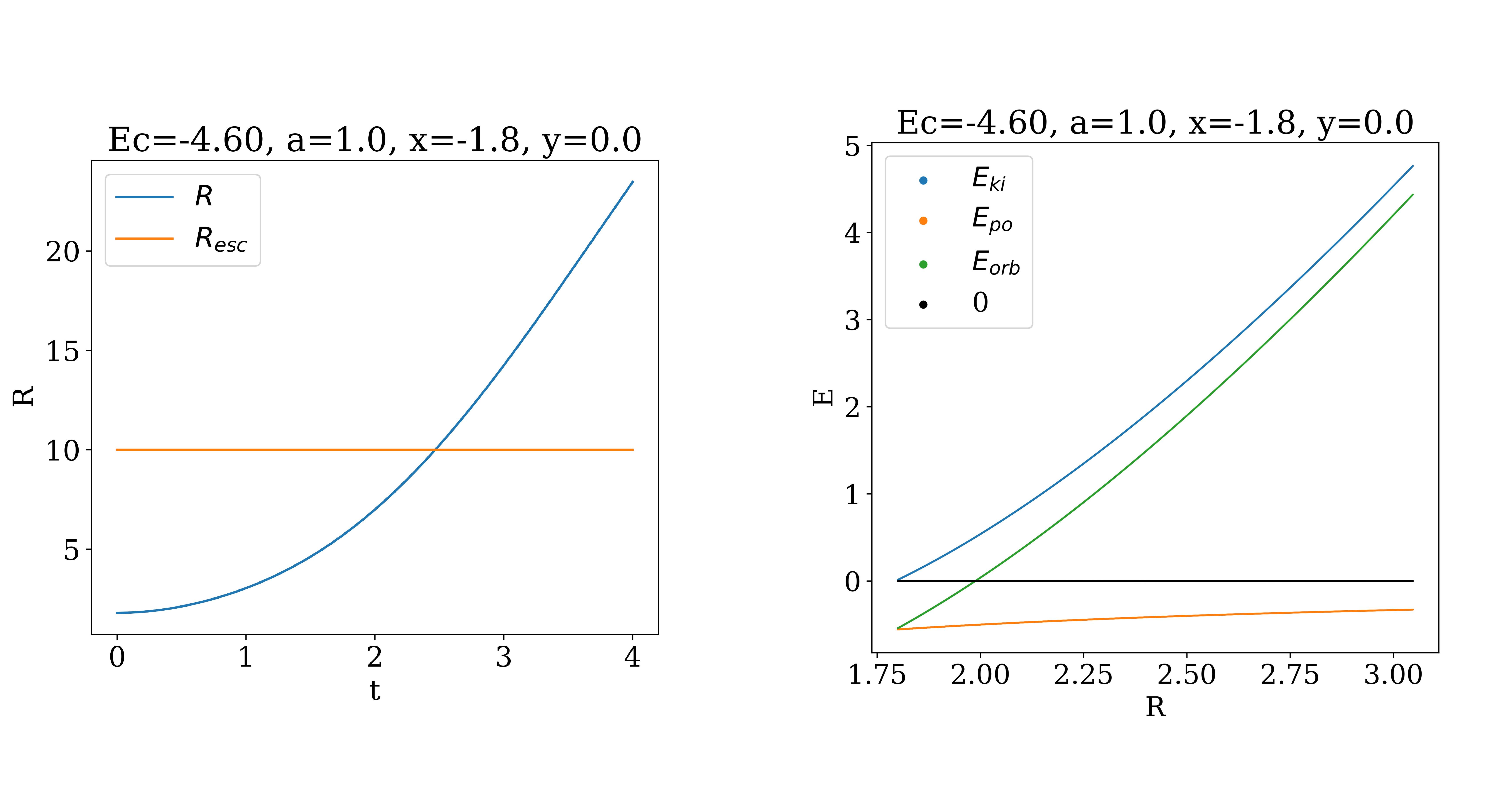}}
\caption{The left panel is the growth of $R$ as a function of time. The horizontal orange line illustrates the escape threshold radius $R_\mathrm{esc}=10$. The right panel is the time evolution of the kinetic energy $E_\mathrm{ki}$ (blue), the potential energy $E_\mathrm{po}$ (orange), and the total orbital energy $E_\mathrm{orb}$ (green). The zero energy level is marked as a black line.  More details are given in the text.}
\label{fig:eorb}
\end{figure*}

\begin{figure}[!t]
\centering
\resizebox{\hsize}{!}{\includegraphics{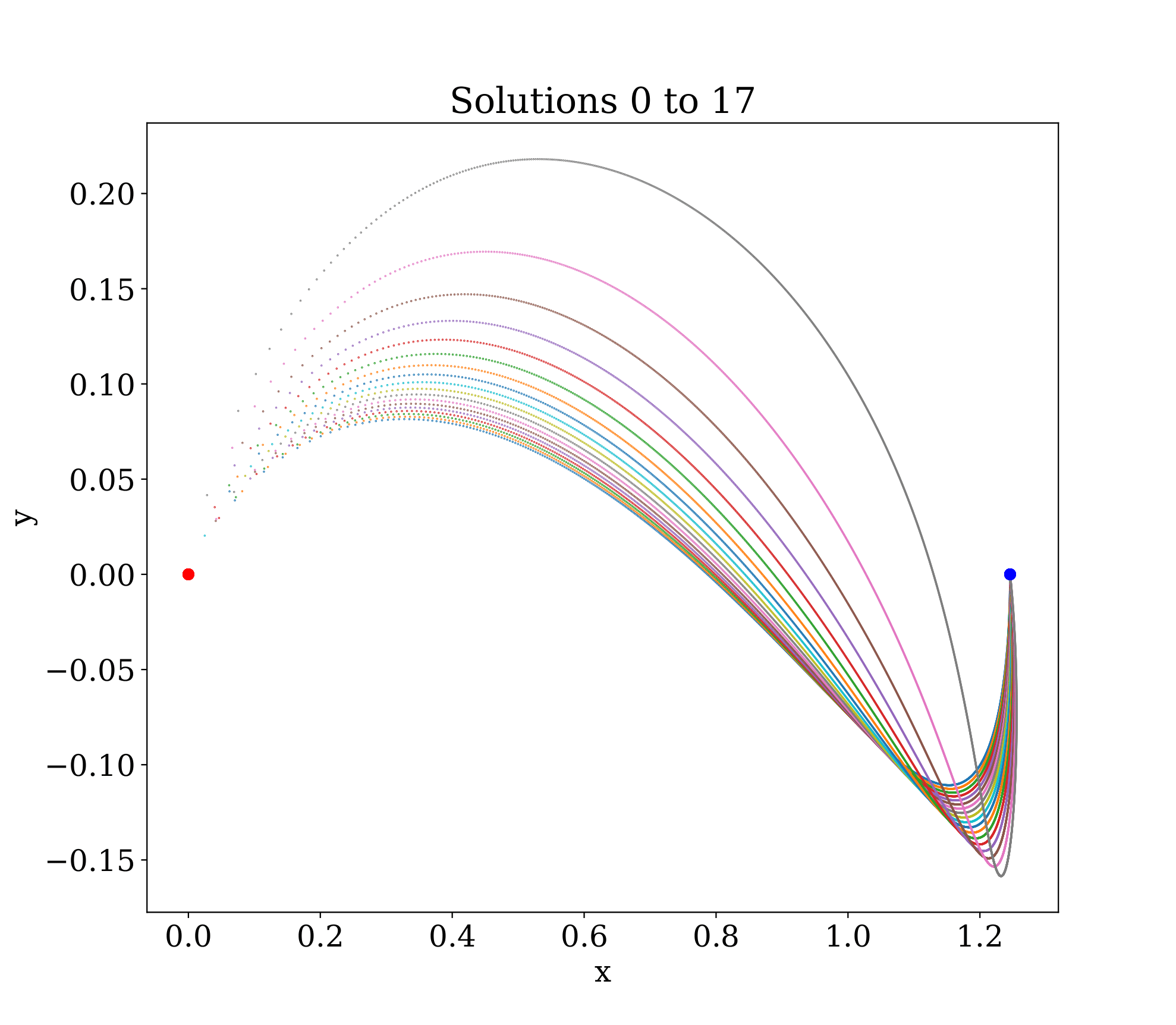}}
\caption{The trajectories for solution 0-17 all collide with the origin. Starting at the blue point and colliding at the red point (the origin).}
\label{fig:col0to17}
\end{figure}

\begin{figure}[!t]
\centering
\resizebox{\hsize}{!}{\includegraphics{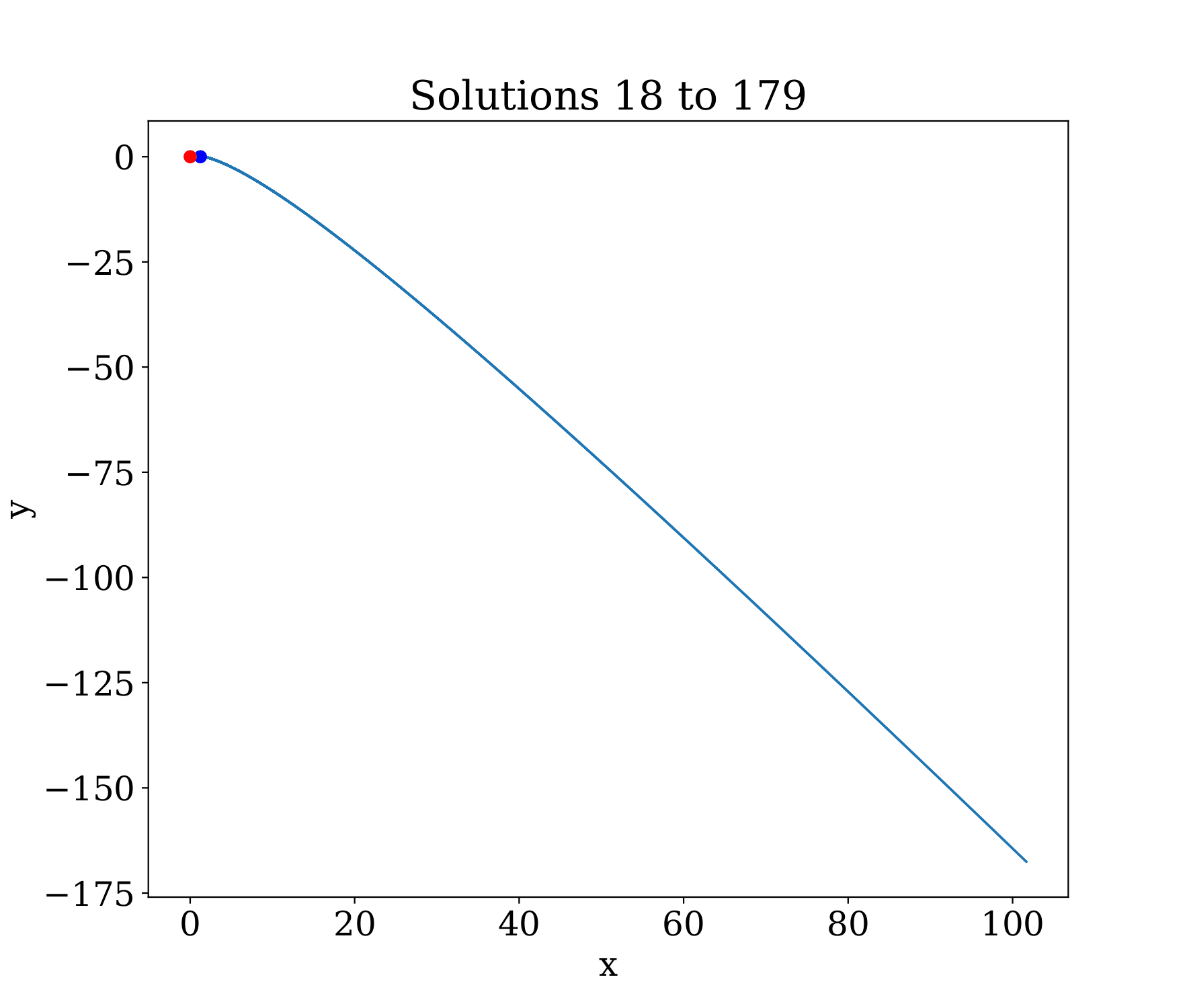}}
\caption{The trajectories for solution 18-179 all look like this; they scatter.}
\label{fig:scat18to179}
\end{figure}

\begin{figure*}[!t]
\centering
\resizebox{\hsize}{!}{\includegraphics{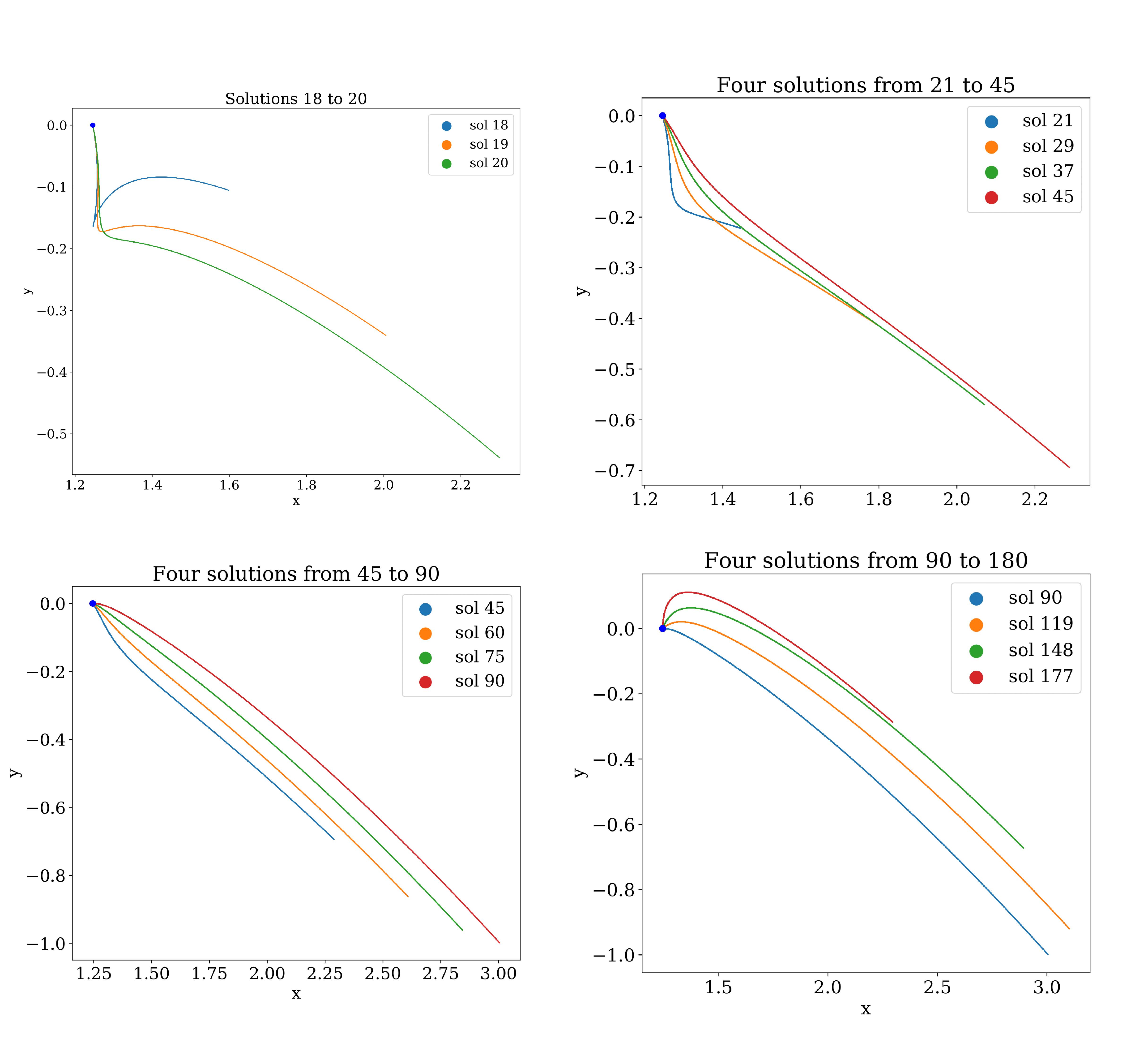}}
\caption{The trajectories zoomed in for solutions from 18 to 179, starting at blue point and scatters.}
\label{fig:scat}
\end{figure*}

\begin{figure}[!t]
\centering
\resizebox{\hsize}{!}{\includegraphics{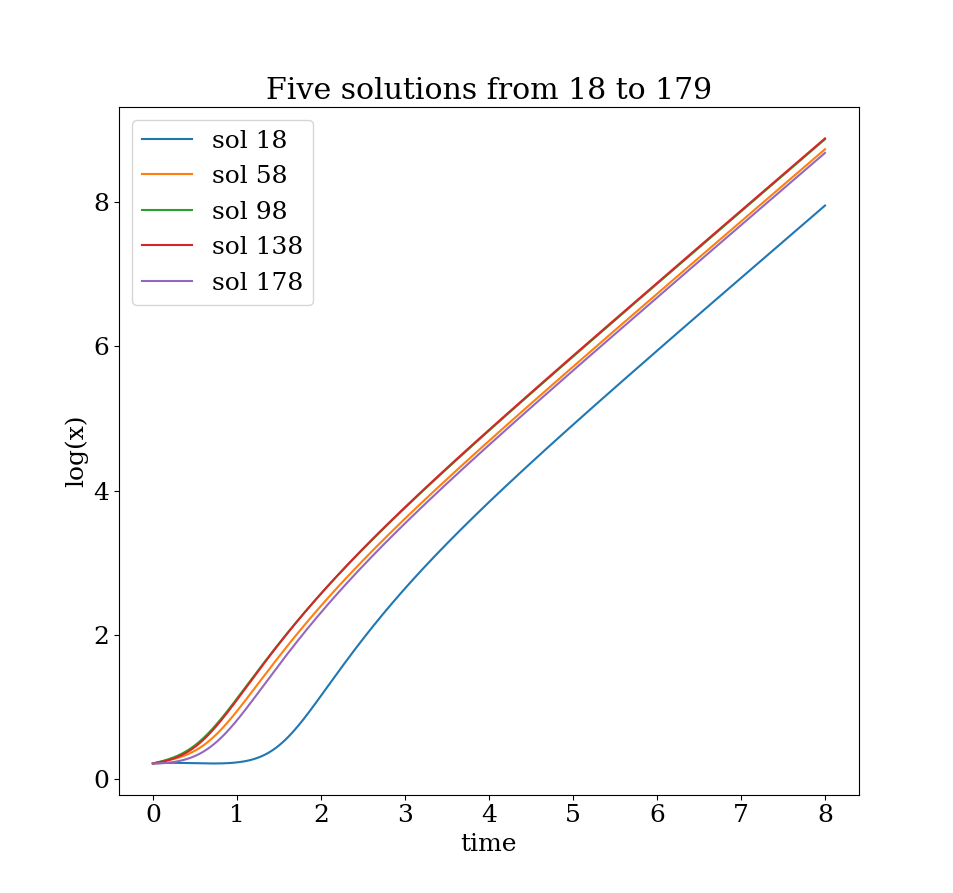}}
\caption{The graph for $\log(x(t))$ for five solutions from 18 to 179. They tend to straight lines after $t=4$.}
\label{fig:expscat}
\end{figure}

\subsection{Above the ground state energy}

In this energy region both bottlenecks are open (cf. Figure \ref{fig:Hillrg}), thus the test particle with initial conditions inside the interior region can escape. Figure \ref{fig:Orabove} presents the orbital structure of the configuration $(x,y)$ space for six values of the gravitational power $\alpha$ and $E_c$ above the ground state energy $E^*$ for each $\alpha$. In the first three diagrams where $\alpha\leq 1.6$ we observe that inside the interior region there are initial conditions with escaping orbit, but the vast majority of the OTD is occupied by initial conditions of collision or bounded orbits. When $\alpha=1.9$ the interior only contains initial conditions with bounded and collision orbits. When $\alpha=3$, the interior region are filled with collision orbits. Again we observe the pattern that as $\alpha$ increases to $2$, the boundaries distinguishing different motions become smoother and remains ``smooth'' for $\alpha\geq 2$.

\subsection{An overview analysis}
The previous color-coded OTDs in the $(x,y)$-plane reveal sufficient information on the phase space, however it is only for a specified energy value. We can use the section $y=\dot{x}=0, \dot{y}>0$, that is, the test particles are launched with these initial conditions so that they start from the $x$-axis with $x=x_0$ and initial velocity pointing vertically upward. Thus we can use $(x, E)$ as coordinates and investigate the behavior of the solutions with various values of the energy $E$. See Figure \ref{fig:xE}.

\subsection{Positivity of orbital energy}
Now in order to verify that the orbits do escape after they exit the disk of radius $R_{\mathrm{esc}}=10$, we follow \cite{Z17} and check the positivity of $E_\mathrm{orb}$. Let $\a=1$ and $E_c=-4.6$, we choose a solution with initial position $x=-1.8, y=0.0$, which is in the escape region. We numerically integrate this solution and we record its distance $R=\sqrt{x^2+y^2}$ from the origin. We choose this position because initially its orbital energy is negative, thus it is interesting to investigate the change of $E_\mathrm{orb}$ from negative to positive with respect to $R$ . Note that for initial position $(x,y)$ where $W(x,y)>0$ and $x>0$ the orbital energy seems to be positive initially, thus there is no need to plot the orbital energy. In the left panel of Figure \ref{fig:eorb} we plot the growth of $R$ as a function of time. One can see that at about $t=2.5$ the trajectory passes the escape threshold radius $R_\mathrm{esc}=10$, and the value of the radius $R$ continues to increase with time. In the right panel of Figure \ref{fig:eorb} we plotted the growth of $E_{\mathrm{orb}}$ as a function of the radius $R$. One can see that the total orbital energy becomes positive around $R=2.0$, which is much lower than $R_\mathrm{esc}=10$. For other energy levels and values of $\alpha$ no orbits whose total orbital energy $E_\mathrm{orb}$ become positive at greater radius that $R_\mathrm{esc}=10$ are found. Therefore, the chosen escape threshold is safe.

Finally, we remark that our computations are restricted to initial conditions with $\dot{y}>0$ and $\dot{x}=0$ for fixed energy. If we take different section of surfaces, we will get different orbital contents in the OTDs. For example, in \cite{Z17} the author took the part $\dot{\varphi}<0$ of the surface section $\dot{r}=0$, where $(r, \varphi)$ is the polar coordinates of $(x,y)$. We choose $\dot{x}=0$ in our computations because it contains more interesting orbital contents. After all, the different section of surfaces in the Hill's problem share the same pattern; that is, as $\alpha$ increases to $2$, the boundaries distinguishing different motions become smoother and remains ``smooth'' for $\alpha\geq 2$.  This is the phenomenon we want to explore in the numerical investigations.

\section{No-return Property and Scattering for the supercritical case}
\label{scat}

For the supercritical Hill's problem, our numerical computations will only be carried out for $\alpha=3$ , as we have similar structures for all $\alpha\geq 2$.

\subsection{Computational methods}

We will only study if there are homoclinic orbits at the Lagrange point $L_2=(q_0, 0)$ where $q_0=\a^{\frac{1}{\a+2}}=3^{\frac{1}{5}}$. The energy will be taken slightly above the ground state energy.

We will fix energy $E=E^*+c$ and initial position at $x(0)=q_0, y(0)=0$, thus the kinetic energy is $E_k:=E-V(q_0,0)=E-E^*=c$. Since $E_k=\frac{1}{2}(\dot{x}^2+\dot{y}^2)$, we take all different directions of the initial velocity, in particular, we take \EQ{\dot{x}(0)=\sqrt{2E_k}\cos(\theta),\quad \dot{y}(0)=\sqrt{2E_k}\sin(\theta),} where $\theta$ ranges from $[-\pi/2, \pi/2]$, i.e. they are shooting outwards in the beginning. If they are shooting inwards, the orbit will collide with the origin as studied in the one-pass theorem in \cite{DI19}.

We take $N$ angles, in Python they are $\theta_i=-\pi/2+\frac{i}{N-1}\pi$, $i=0,\cdots, N-1$. For each initial condition $$\Gamma_i=(q_0,0,\sqrt{2E_k}\cos(\theta_i), \sqrt{2E_k}\sin(\theta_i)),$$ we integrate the equation (\ref{eq:hlp}) and plot their trajectories in the $(x, y)$ plane.

\subsection{No-return: collision or scattering}

We take $\a=3$, and fix $E=-6.4\approx E^*+0.066$, $N=180$. The solution with initial condition $\Gamma_i$ is called ``solution $i$'' for $i=0, \cdots, 179$. In Figure \ref{fig:col0to17}, we see solutions from $0$ to $17$ all collide with the origin. The blue dot is the starting position, i.e. $L_2$, and the red dot is the origin. It is interesting to note that these trajectories collide with the origin even though initially they are not pointing towards the origin. In Figure \ref{fig:scat18to179} we see all solutions from $18$ to $179$ scatter (or escape).

For the scattering solutions, we also zoomed in (decrease the maximal integration time) to investigate their behavior in the beginning. See Figure \ref{fig:scat}. We see there are turnings for solution $18$ to about $21$, and for solution $21$ to about $90$ they escape to infinity monotonically (without turnings), and for solutions from $91$ to $179$ there are turnings. After all, they escape to infinity. We conclude that there are no ``homoclinic'' returns at the Lagrange point $L_2$; the orbits shooting outwards initially either collide with the origin or escape to infinity.

Moreover, for the escaping solutions 18 to 179, we plot the graph $\log(x(t))$, see Figure \ref{fig:expscat}. After some time, say $t=4$, the graph of $(t, \log(x))$ tends to be a straight line, indicating that they scatter exponentially.

\section{Conclusions}
\label{conc}

In this paper we have numerically investigated the Hill's lunar problem with homogeneous potential. The main goal of the paper was to show the orbital difference between the subcritical ($\alpha<2$) and supercritical ($\alpha\geq2$) cases for the Hill's problem. Another attempt was to support the conjecture of the no-return property in the outer region for the supercritical case, as well as to show the generic exponential scattering for global solutions.

For the orbital classification, we have chosen several values of the power of the potential and examined the two most interesting Hill' regions, i.e. below the ground state energy and above the ground state energy. Each of the Hill's region, we used uniform grids of $1024\times1024$ initial conditions that are regularly distributed on the configuration $(x,y)$-plane with $\dot{y}>0$. All orbits were integrated with initial conditions inside the region $-2\leq x,y\leq 2$.

For the no-return property and scattering for the supercritical case, we chose $\alpha=3$ and the Hill's region with energy level above the ground state. We investigated the orbits when they are launched at the right Lagrange point and with initial velocity pointing towards the exterior part of the Hill's region.

The main numerical results in the paper can be summarized as follows:

\begin{enumerate}
\item When the energy is below the ground state, we see that in the bounded Hill's region, both collision motion and bound motion are presented for $\alpha<2$. While for $\alpha\geq 2$, there are only collision orbits inside.
\item When the energy is above the ground state, we see that in the interior region ($W<0$), both collision motion, bound motion and escaping motion are possible for $\alpha<2$. While for $\alpha\geq 2$, there are only collision orbits inside.
\item In both types of the configuration plane, the ``fractal" nature of the plane was reduced (or disappeared) as the exponent $\alpha$ increased and exceeded $2$.
\item We also showed that near the Lagrange point $L_2$, there are no returning orbits. Moreover, the escaping orbits for $\alpha>2$ are mainly exponentially scattering.

\end{enumerate}

Based on these results, we may say that this numerical investigation serves as a nice illustration and support for the analytical results \cite{DI19,DIN19} on the problem. This work also encourages us to study the strong force restricted three-body problem, for which the Hill's lunar problem is just a limiting case. It is our hope to fully investigate the significant dynamical differences between the strong force gravitational field and the weak (e.g. Newtonian) gravitational field.


\begin{thebibliography}{}
\footnotesize

\bibitem{AM08} R. Abraham, J. Marsden: Foundations of Mechanics. Second Edition, Amer. Math. Soc. (2008)

\bibitem{MO17} K.~R. Meyer, D.~C. Offin: Introduction to Hamiltonian Dynamical Systems and the N-Body Problem. Third Edition Applied Mathematical Sciences Springer (2017)

\bibitem{MS95} J.~K. Moser, C.~L. Siegel: Lectures on Celestial Mechanics. Classics in Mathematics Springer (1995)

\bibitem{G75} W.~B. Gordon: Conservative dynamical systems involving strong forces. Trans. A. M. S., \textbf{204}, 113-135 (1975)

\bibitem{MG81} R. McGehee: Double collisions for a classical particle system with nongravitational interactions. Comment. Math. Helvetici \textbf{56}, 524-557 (1981)

\bibitem{DI20} Y. Deng, S. Ibrahim: Global Existence and Singularity of the N-body Problem with Strong Force. Qual. Theory Dyn. Syst. \textbf{19}, 49 (2020)

\bibitem{DI19} Y. Deng, S. Ibrahim: Global existence and singularity of the Hill's type lunar problem with strong potential. \url{https://arxiv.org/abs/2010.05130} (2020)



\bibitem{NaSc12} K. Nakanishi, W. Schlag: Global dynamics above the ground state energy for the cubic NLS equation in 3D. Calc. Var. and PDE \textbf{44}, 1-45 (2012) 


\bibitem{AkIb18} T. Akahori, S. Ibrahim, H. Kikuchi, H. Nawa: Global dynamics above the ground state energy for the combined power-type nonlinear Schr\"odinger equations with energy-critical growth at low frequencies. (To appear in Memoirs of the A.M.S.)



\bibitem{Z17} E. Zotos: Investigating the planar circular restricted three-body problem with strong gravitational field. Meccanica \textbf{52}, 1995-2021 (2017)

\bibitem{BTS96} L. Benet, D. Trautman, T. Seligman: Chaotic scattering in the restricted three-body problem I. The Copenhagen problem. Celest. Mech. Dyn. Astron. \textbf{66}, 203-228 (1996)

\bibitem{DIN19} Y. Deng, S. Ibrahim, K. Nakanishi: Dynamics classification for the supercritical Hill problem. (in preparation)

\bibitem{WBW05} H. Waalkens, A. Burbanks, S. Wiggins: Escape from planetary neighborhoods. Mon. Not. R. Astron. Soc. \textbf{361}, 763-775 (2005)

\bibitem{MS82} K.~R. Meyer, D.~S. Schmidt: Hill's Lunar Equations and the Three-Body Problem. J. Diff. Eq. \textbf{44}, 263-272 (1982)

\bibitem{AVS01} J. Aguirre, J.~C. Vallego, M.~A.~F. Sanju\'{a}n:Wada basins and chaotic invariant sets in the H\'{e}non-Heiles system. Phys. Rev. E \textbf{64}, 066208 (2001)

\bibitem{AVS09} J. Aguirre, R.~L. Viana, M.~A.~F. Sanju\'{a}n: Fractal structures in nonlinear dynamics. Rev. Mod. Phys. \textbf{81}, 333-386 (2009)

\bibitem{DWGGS16} A. Daza, A., Wagemakers, B., Georgeot, D., Gu\'{e}ry-Odelin, M.~A.~F, Sanju\'{a}n: Basin entropy: a new tool to analyze uncertainty in dynamical systems. Scientific Reports \textbf{6}, article number: 31416 (2016)

\bibitem{S01} C. Skokos: Alignment indices: A new, simple method for determining the ordered or chaotic nature of orbits. J. Phys. A: Math. Gen. \textbf{34}, 10029-10043 (2001)

\bibitem{PTVF92} H.~P. Press, S.~A. Teukolsky, W.~T., Vetterling, B.~P. Flannery: Numerical Recipes in FORTRAN 77, 2nd Ed., Cambridge Univ. Press, Cambridge, USA (1992)
    
\bibitem{W03} S. Wolfram: The Mathematica Book. Wolfram Media, Champaign (2003)

\bibitem{N04} J. Nagler: Crash test for the Copenhagen problem. Phys. Rev. E \textbf{69}, 066218 (2004)

\bibitem{N05} J. Nagler: Crash test for the restricted three-body problem. Phys. Rev. E \textbf{71}, 026227 (2005)

\end{thebibliography}
\end{document}